\documentstyle[aps,prbbib]{revtex}
\begin{document}
\draft
\title{Quantum Hall Effects at Finite Temperatures}
\author{Sudhansu S. Mandal,$^\ast$
S. Ramaswamy,$^\dagger$
and V. Ravishankar$^\ast$}
\address{$^\ast$Department of Physics,
Indian Institute of technology, \\
	  Kanpur 208 016, INDIA \\
     $^\dagger$Mehta Research
   Institute of Mathematics and Mathematical Physics, \\
	 10, Kasturba Gandhi Marg, Allahabad 211 002, INDIA}

\maketitle

\begin{abstract}
We study the finite temperature (FT) effects on integer
quantum Hall effect (IQHE) and fractional quantum Hall effect (FQHE)
as predicted by the composite fermion model.
We find that at $T\neq 0$, universality is lost, as is quantization
because of a new scale $T_0=\pi\rho /m^\ast p$. We find
that this loss is not
inconsistent with the experimentally observed accuracies. While the
model seems to work very well for IQHE, it agrees with the bulk results
of FQHE but is shown to require refinement in its account
of microscopic properties such as the effective mass.
Our analysis also gives a qualitative account of the threshold
temperatures at which the FQHE states are seen experimentally.
Finally, we extract model
independent features of quantum Hall effect at FT, common to all
Chern-Simons theories that employ mean field ansatz.
\end{abstract}


 \medskip

%\newpage

The purpose of this paper is to study quantum Hall effect (QHE), both
integer (IQHE) \cite{klit} and fractional (FQHE), \cite{tsui1}
in the compsite fermion model (CFM) at finite temperatures.
The merit of such a study is recognized if we recall
that IQHE is used to measure the fine structure constant -- to a great
accuracy of 0.01 ppm, \cite{cage1} -- with a complex `dirty system' and
yet without any dependence on the sample properties such as the shape,
density
and mass of electron etc. It is pertinent to investigate upto what
temperature  this universality and the precise quantization of the Hall
resistivity $\rho_{xy}$ will survive, given a required
degree of accuracy.
Similarly, the associated phenomenon, FQHE, also exhibits
quantization \cite{tsui1} at fractional filling factors; and as it was
emphasized by Laughlin \cite{laugh1} long back, the experimentalists
\cite{expts} have found it necessary to go to lower temperatures (and
higher magnetic fields) to see more and more plateaus.
This fact has to be
accounted for, not merely qualitatively, and it is of interest
to see if the CFM, which appears to be a valid description
at $T=0$, is also reliable at finite temperatures.

We point out that there are two questions here -- the dependence of
quantization, and the dependence of plateau width on temperature. There
is a wealth of experimental data available (although a
lot more are needed
for a full study) which we shall discuss contextually below.
In this paper,
we merely address the first question since for simplicity, we choose a
pure system -- which yields a vanishing diagonal
resistivity $\rho_{xx}$,
and also does not incorporate the plateau widths.
These drawbacks will be
remedied in a more detailed presentation of our work.

There has been a considerable gain in our understanding
of QHE, thanks to
the contributions made by Ando et al., \cite{ando}
Prange, \cite{prang} Laughlin,
\cite{laugh2}
Halperin, \cite{halp1} and Streda \cite{streda}
for IQHE and by the seminal work of Laughlin \cite{laugh3} which was
followed by that of Haldane \cite{hald} and Halperin. \cite{halp2}
Halperin \cite{halp2} was the first to recognize the
anyonic nature of the
Laughlin's wave function which suggests a Chern-Simons (CS) interaction
in the system. This idea has been employed by Jain \cite{jain} in
his CFM where, however, the CS term does not
transmute the statistics even while playing a dynamical
role. A field theoretic version of the Jain picture has been
given by
Lopez and Fradkin \cite{lopez}.

CFM has the attractive feature of treating IQHE and FQHE on an equal
footing (and is thus a convenient model to study
the finite temperature (FT)
properties, atleast in a first approach \cite{fnote} ).
Indeed, when the system is subjected to an external magnetic field $B$,
a part $b$ of $B$ gets
attached as singular flux tubes to the electrons and
becomes the dynamical
magnetic field in the CS action. The Hall resistivities are then derived
by proceeding with a mean field (MF) ansatz which smears
the flux lines to a
uniform value of $b$. A study of the fluctuations
about this MF state then
immediately yields $\rho_{xy}$. We propose to extend this study to
FT below. Finally, we shall argue that it is possible to
extract those features that are common to all CS based theories which
invoke the MF ansatz of which the CFM is but an instance.

Consider then a system of nonrelativistic spinless fermions
in an external
magnetic field of strength $B$ confined to
the direction perpendicular to
the plane of the system. The Jain proposal
\cite{jain} consists of introducing
an internal CS magnetic field of strength $b$ such
that the system sees an
effective field $B_{\mbox{eff}}=\vert B+b\vert $. The next step is to
attribute a strength $\mp (2\pi\rho /e)(2s)$
to the CS field and a strength
$(2\pi\rho /e)(1/p)$ to $B_{\mbox{eff}}$,
where $e$, and $\rho$ are the charge
and density of the fermions respectively, and $s$ and $p$ are
integers. This assignment immediately leads
to a state of filling fraction
$\nu =p/(2sp\pm 1)$. The study of such an
effective system is accomplished
through the following Lagrangian density,
\begin{eqnarray}
{\cal L} &=& \psi^\ast iD_0\psi -\frac{1}{2m^\ast} \left\vert D_k \psi
\right\vert^2 +\frac{\theta}{2} \epsilon^{\nu \lambda \rho}a_\nu
\partial_\lambda a_\rho +\psi^\ast \mu \psi  \nonumber \\
  & & -eA_0^{\mbox{in}} \rho
+\frac{1}{2} \int d^3x^\prime \, A_0^{\mbox{in}} (x) V^{-1}
(x-x^\prime) A_0^{\mbox{in}} (x^\prime)    \; .
\label{eq1}
\end{eqnarray}
Here $D_\nu =\partial_\nu -ie(a_\nu+A_\nu+A_\nu^{\mbox{in}} )$
$(a_\nu \,
,\, A_\nu$, and $A_\nu^{\mbox{in}}$ being the CS, external and internal
Maxwell gauge fields respectively),
$\mu$ is the chemical potential, $m^\ast$ is the effective mass of
the fermions, and
$\theta= \pm (e^2/2\pi)(1/2s)$ is the CS parameter.
Finally, $V^{-1} (x-x^\prime
)$ represents the inverse of the instantaneous charge charge
interaction potential (in
the operator sense). \cite{lopez}
The above Lagrangian density is equivalent to the usual four fermion
interaction term as is considered by Lopez and Fradkin. \cite{lopez}
Observe that the IQHE corresponds to the choice
$s=0$ (i.e., $\theta =\infty $) which implies
a net mean zero value for the CS
field. FQHE follows from the choice $s\neq 0$,
i.e., $\langle b \rangle \neq 0$.
Note also that the electrons interact with each other via $1/r$
or some other short range potential, i.e., the internal dynamics is
governed by $(3+1)$ dimensional Maxwell Lagrangian as is appropriate for
the medium.

The procedure for evaluating the FT properties of the system with the
above Lagrangian density is standard. We do not discuss the details here
since they have been presented in the allied context of Chern-Simons
superconductivity elegantly by Randjbar-Daemi et. al., \cite{randj} and
has been extensively used. \cite{ssman} In brief, we construct the
partition function ($\beta =1/T$ being the inverse temperature),
\begin{equation}
{\cal Z} = \int [da][dA][dA_0^{\mbox{in}}][d\psi ]
[d\psi^\ast ]\, e^{-\int_0^\beta
d\tau \int d^2{\bf r} {\cal L}^{(E)}} \; ,
\label{eq2}
\end{equation}
which on integration over the fermionic fields, in the MF ansatz,
factors
into ${\cal Z}={\cal Z}_{MF} {\cal Z}_f $ after the usual saddle point
computation. Here ${\cal L}^{(E)}$ is the euclidian version of
${\cal L}$ (\ref{eq1}).
The MF part of the partition function is given by
$(1/A)\ln\,{\cal Z}_{MF} = \rho_l \sum_{n=0}^\infty
\sum_{j=-\infty}^\infty \ln\, \left[ \epsilon_n-\mu
+i\omega_j \right]$,
where $\epsilon_n=(n+1/2)\omega_c $,
($\omega_c=(e/m^\ast)B_{\mbox{eff}}$
being the effective cyclotron frequency), is the energy corresponding to
$n$ th Landau level, $\omega_j=(2j+1)\pi/\beta$ is the Matsubara
frequency, $\rho_l=m^\ast \omega_c/2\pi $
is the degeneracy per unit area
in each level, and $A$ is the area of the system. The corresponding
thermodynamic potential is obtained as
$(\Omega /A)=-(\rho_l /\beta)
\sum_{n=0}^\infty \ln \left( 1+\exp [-\beta
(\epsilon_n-\mu)] \right)$,
from which all the MF properties can be inferred.

Writing the fluctuating part of the partition function as
${\cal Z}_f = \int [da][dA][dA_0^{\mbox{in}}]\,\exp
[-S_{\mbox{eff}}]$,
(where we have expanded upto second order
in the gauge field fluctuations
around the MF configuration), we identify $S_{\mbox{eff}}$ with the
one-loop effective action which is given by
\begin{eqnarray}
S_{\mbox{eff}} &=& -i\frac{\theta}{2}\epsilon^{\nu
\lambda \rho}\int d^3x \,
a_\nu \partial_\lambda a_\rho +\frac{1}{2}\int d^3x\int d^3x^\prime
(a_\mu+A_\mu+A_\mu^{\mbox{in}}\delta_{\mu 0} ) \nonumber  \\
 & & \times \Pi^{\mu \nu} (x\, ,\,
x^\prime) (a_\nu+A_\nu+A_\nu^{\mbox{in}}
\delta_{\nu 0} ) -\frac{1}{2}\int d^3x
\int d^3x^\prime \, A_0^{\mbox{in}}
(x) V^{-1} (x-x^\prime)
A_0^{\mbox{in}} (x^\prime) \; .
\label{eq6}
\end{eqnarray}
The current correlation functions $\Pi^{\mu \nu }
(x,x^\prime)\equiv \delta
\langle j^\mu (x) \rangle /\delta {\cal A}_\nu (x^\prime)$,
where $j^\mu $ is the fermionic current,
and ${\cal A}_\nu $ is the sum of
all the gauge fields, have to be determined.
Using Galilean and gauge invariance, we write (in the momentum space)
\begin{eqnarray}
\Pi^{\mu \nu} (\omega \, ,\, {\bf q}) &=& \Pi_0 (\omega\, ,\, {\bf q})
(q^2g^{\mu \nu} -q^\mu q^\nu )+(\Pi_2-\Pi_0) (\omega \, ,\, {\bf q})
\nonumber  \\
& & \times ({\bf q}^2 \delta^{ij}-q^iq^j)\delta^{\mu i}\delta^{\nu j}
+i\Pi_1 (\omega \, ,\, {\bf q}) \epsilon^{\mu \nu \lambda}q_\lambda \; ,
\label{eq7}
\end{eqnarray}
with $\Pi_0=\bar{\Pi}_0+\Gamma/{\bf q}^2 $.
In the low $q$ limit, we find the form factors to be
\begin{eqnarray}
\bar{\Pi}_0 &=& \frac{e^2}{2\pi\omega_c}\sum_{n=0}^\infty f_n
-\frac{e^2\beta}{16\pi} \sum_{n=0}^\infty (2n+1) \mbox{sech}^2
\frac{\beta}{2} \Omega_n  \; , \;
\Gamma = \frac{e^2 m^\ast \beta \omega_c}{8\pi}\sum_{n=0}^\infty
\mbox{sech}^2 \frac{\beta}{2} \Omega_n  \; , \nonumber \\
\Pi_1 &=& \bar{\Pi}_0 \omega_c   \; , \;
\Pi_2 = \frac{e^2}{2\pi m^\ast }\sum_{n=0}^\infty (2n+1)f_n
-\frac{e^2\beta\omega_c}{32\pi m^\ast }\sum_{n=0}^\infty (2n+1)^2
\mbox{sech}^2 \frac{\beta}{2} \Omega_n   \;  ,
\label{eq8}
\end{eqnarray}
where $f_n=\left[ 1+\exp \left( \beta \Omega_n \right) \right]^{-1}$ and
$\Omega_n =\epsilon_n -\mu $.
Note that the exclusively thermal form factor
$\Gamma$ has the interesting
property \cite{ssman} that $\Gamma =0$ for ${\bf q}^2=0$, $\omega
\rightarrow 0$ and it is nonzero for $\omega =0$, ${\bf q}^2 \rightarrow
0$. The FT properties of the system are driven by
the temperature behaviour
of these form factors. In particular, the parity and time reversal
violation caused by the external magnetic field acts through the form
factor $\Pi_1$, which indeed controls the behaviour of $\rho_{xy}$ with
changing temperature.

The experiments are performed at temperatures
in the range from 20 mK to a
few K. We need to evaluate the form factors in this regime. If we are
interested in extremely small deviations from the zero temperature value
or in estimating whether the lowest temperatures
reached or small enough,
a low temperature (LT) expansion of the form factors should suffice. In
that case, they are
analytically evaluated as a perturbation in $\exp [-\beta\omega_c /2 ]$
(see Ref.~18 and 19 for details of calculation) and are found to be
\begin{eqnarray}
\bar{\Pi}_0 &=& \frac{e^2 m^\ast p^2}{4\pi^2\rho} (1-4y) \; , \;
\Gamma = 4\frac{e^2 m^\ast}{\pi}y  \; ,  \nonumber \\
\Pi_1 &=& \frac{e^2p}{2\pi} (1-4y) \; , \;
\Pi_2 = \frac{e^2p^2}{2\pi m^\ast} (1-4y) \; ,
\label{eq9}
\end{eqnarray}
where $y\equiv (T_0/T)\exp [-T_0/T]$ with $T_0=\pi\rho /m^\ast p$.
At higher temperatures, we need the exact values which can only be
obtained numerically.

Given the form factors, a straight forward
linear response analysis which
involves (see Ref.~19 for procedure) the average over the internal
fluctuations as well as a coupling to a weak external electric field,
yields the Hall resistivity to be
\begin{equation}
\rho_{xy}(\omega \, ,\, {\bf q}) = \frac{\Pi_0(\Pi_0\omega^2 -\Pi_2 {\bf
q}^2) -(\Pi_1+\theta)^2 -\Pi_0\theta V(q){\bf q}^2}{\Pi_0\theta
(\Pi_0\omega^2 -\Pi_2 {\bf q}^2 ) -\Pi_1\theta (\Pi_1+\theta) } \; .
\label{eq10}
\end{equation}
Note that the diagonal resistivity vanishes
by virtue of the purity of the
system. For the same reason, the quantizations
occur at specific values of
$B$ which are recognized to be the central values of the plateaus seen
experimentally. As remarked earlier, we concentrate on
$\rho_{xy} (T)$ at this central value.

Before we present our results,
and compare them with experiments
wherever possible, it should be noted that the behaviour of the form
factors, and hence the response functions,
has a crucial dependence on the
choice of $V(q)$. Recall that the Laughlin wave function which correctly
describes the states with filling fractions given by $\nu = 1/(2k+1)$,
$k=1,2,\cdots$, has been shown numerically by Haldane \cite{hald}
to be exact for a large class of short
range repulsive potentials.
It is clear from our FT analysis by Eqs. (\ref{eq7}) --
(\ref{eq10}) that if the static conductivity $\sigma^s
\equiv -1/\rho^s$, where $\rho^s
\equiv \rho_{xy} (\omega =0 \, ,\, {\bf q}^2
\rightarrow 0)$, is to survive,
then we require $V(q)\rightarrow C $ (const.)
as ${\bf q}^2 \rightarrow 0$. If $V(q)$
diverges as ${\bf q}^2 \rightarrow 0$ as it could happen for potentials
which are long ranged, i.e.,
$V(r)\rightarrow 0$ as $r\rightarrow \infty$
slower than $1/r^2$, then $\sigma^s$ would
have its support only at $T=0$.
Clearly, such interactions are ruled out
from this analysis. Further $V(q)
\rightarrow C\neq 0$ (as it would happen for $V(r)\sim 1/r^2$ or $\delta
({\bf r})$ ) is a threshold case in the
sense that $\sigma^s (T\neq 0)\neq
0$, but is sensitive to the strength of the interaction. This means the
universality has its support only at $T=0$ with a strong dependence on
strength at $T\neq 0$, which is again unphysical. We
conclude that $V(r)$ should be more short
ranged, in confirmation with the
analysis of Haldane \cite{hald} and also Trugman
and Kivelson \cite{trug} who showed the
exactness of Laughlin's wave function for one
such potential. Indeed,
$\sigma^s$ is then independent of $V(q)$ as it should be by continuity
requirement.

Note that if we define $\rho^d \equiv \rho_{xy}(\omega
\rightarrow 0\, , \, {\bf q}^2 =0)$, $\rho^s \neq \rho^d $.
It has a different temperature evolution, but is again governed
by $T_0$, which we shall discuss in the detailed paper.

In order to discuss the limit on accuracy of quantization imposed by
temperature, consider $\rho^s$ at very small values of $y$. It has the
analytic form
\begin{equation}
\rho^s (T)=\rho^s (0) + 4(2\pi/e^2p)y \left[ 1- \frac{4sp}{2sp\pm 1}
\right]  \; ,
\label{eq11}
\end{equation}
with $\rho^s (0)=(2\pi /e^2 )(2sp\pm 1)/p $.
 From the expression (\ref{eq11}), it is clear that the temperature
dependence is indeed accompanied by a corresponding deviation from
universality in virtue of its dependence on the parameter $T_0$
which is the only sample specific parameter that enters the analysis.
The argument is more robust.
Indeed, at any temperature, although we can not evaluate $\rho^s (T)$
analytically, it is easy to check that the Hall defect
\begin{equation}
{\cal R} \equiv \left\vert \frac{\rho^s (T)
-\rho^s (0)}{\rho^s (0)} \right\vert
\label{eq12}
\end{equation}
is a function of the dimensionless variable $T_0/T$. This kind of
dependence and the specific form of $T_0$ is a
reflection of the MF ansatz which
introduces the fundamental scale $\omega_c$, the cyclotron frequency.

Having discussed these general features, we now consider IQHE in
more detail. This is of great significance since at $T=1.8$ K, the fine
structure constant was measured to an accuracy
of 5 ppm by von Klitzing et
al. \cite{klit} A further lowering of the temperature has led to an
accuracy of 0.01 ppm suggesting that the universality is achieved
asymptotically as $\beta \rightarrow \infty$.

Fig.~1 shows how ${\cal R}_0$ ( which is
simultaneously a measure of both
universality and quantization loss) evolves with temperature for an
accuracy range 0.01 ppm to $1\%$ for $\nu =1$.
In fact, we find that if we fix the
value of ${\cal R}_0$, the temperature
$T_{{\cal R}_0}$ at which ${\cal R}_0$ is achieved
follows a simple expression
\begin{equation}
\frac{\rho}{m^\ast pT_{{\cal R}_0}} = C^\prime  \; ,
\label{eq13}
\end{equation}
where the const. $C^\prime$ depends only on ${\cal R}_0$.
For example, the values of
$C^\prime$ at ${\cal R}_0=10^{-n}$ are approximately given by
0.325+0.776(n+1) for $3\leq n\leq 8$.
It would be extremely interesting if
some or all of these predicted features can be verified experimentally.
Any confirmation of a dependence of $T_{{\cal R}_0}$ on
$\rho /m^\ast $ or of scaling with
$p$ would be a striking vindication of
the MF ansatz. In any case, we note
that a requirement of higher accuracy at any given temperature demands
correspondingly higher values of $\rho /m^\ast $. Further, it is also
clear that the temperature scales with $\rho /m^\ast$ for a fixed ${\cal
R}_0$ and $p$.

Yoshihiro et al. \cite{yoshi}
report that $\Delta \rho^s =\rho^s (T)-\rho^s (0)$ has the form
$c\rho_{xx}^{\mbox{min}}
$ (with $c=-0.1$) over a temperature range 0.5 K -- 1.6 K, and for the
densities in the range 1 -- 3$\times 10^{12}$ cm$^{-2}$. We are not in a
position to make any comparison with this experimental observation.
However, our analysis shows that $\Delta \rho^s \neq 0$ even if
$\rho_{xx}=0$.
Further at LT, the accuracy
can be expressed in a simple form ${\cal R}_0 =4y$ which is obtained by
putting $s=0$ in Eqs.
(\ref{eq11} and \ref{eq12}), as is appropriate here.
The measured accuracy of 0.2 ppm by Yoshihiro
et al. \cite{yoshi} at the lowest temperature mentioned above, i.e., at
$T=0.5$ K in the above range of densities
and for a narrow range of $B=9$
--10.5 T for the states $p=4,$ 8 and 12
is consistent in our analysis as it yields a reasonable value of
$m^\ast \simeq 0.7$ $m_e$.

However, we have a later measurement of $\rho^s (T)$ by Cage et al.
\cite{cage2} at $T=3$ K and $1.2$ K.
The corresponding experimental values
of ${\cal R}_0$ are 4.2 ppm and 0.017 ppm for $p=4$. The values of $\rho
/m^\ast$ may again be obtained using Eq.
(\ref{eq13}) and they turn out to
be 63.3 K and 34.2 K respectively. The rather significant difference in
the value of $\rho /m^\ast $ obtained possibly indicates that
the role of
impurities becomes more important
(as indeed the experimentalists find) at
such relatively higher temperatures, especially since $\rho_{xx}$ has a
strong dependence on $T$.

Finally, before we go on to discuss FQHE, we observe that
the experimental results can be used to place
an upper limit on the value of $\rho /m^\ast p$ within the model
considered here. These are summarized in Table~I.

We now discuss FQHE for which there are
some data available \cite{tsui1,storm}
on the slope
of the plateau.
Both $\rho_{xx} (T)$ and $\rho_{xy} (T)$ have also been
measured, the former primarily for the purpose of extracting the gap
energy $E_g$. Theoretically, the temperature dependence of ${\cal R}$ in
this case is given by Eqs. (\ref{eq11}
and \ref{eq12}) as a functions of
$s$ and $p$. Similar to (\ref{eq13}), for a fixed ${\cal R}_\pm$,
the scaling is now
generalized to
\begin{equation}
\frac{T_0}{T}e^{-T_0 /T} = \left\vert
\frac{{\cal R}_\pm}{4} \frac{(2sp\pm 1)^2}{-2sp\pm 1} \right\vert
  \; ,
\label{eq14}
\end{equation}
where ${\cal R}_+=(2sp-1)4y/(2sp +1)^2$ and ${\cal R}_-=4y(2sp+1)
/(2sp -1)^2$ are
the accuracies corresponding to the states with
antiparllel (`$+$') and parallel (`$-$')
flux attachments. Clearly, for a fixed $s$, and $p$,
`$+$' states will be seen with
larger accuracy than `$-$' states at the same temperature
for a given sample. For example,
$\nu =3/7$ state should be seen at a
higher accuracy than $\nu =3/5$ state,
which is indeed true as it has been seen experimentally. \cite{chang1}
Fig.~1 shows how the temperatures $T_{{\cal R}_\pm}$, (for which
the accuracies are ${\cal R}_\pm $) vary with ${\cal R}_\pm $
for $(2,1)_+$ and $(2,1)_- $ states, where
we have used the notation $(p,s)_\pm $ to denote the states.

Chang et al. \cite{chang2} report that at 65 mK and for $\rho =
2.1 \times 10^{11}$ cm.$^{-2}$, the quantization at
$\nu=5/3$ has an accuracy of 1.1 parts in
$10^3$ and at $\nu =2/3$, it has
an improved accuracy of 3 parts in $10^4$.
Our theoretical estimate at $\nu =2/3$
($(2,1)_-$) yields
$m^\ast  =3.9 \; m_e$.
Further, the estimates of $m^\ast$ from CFM for different
filling fractions using the results (measured at $T=90$ mK) obtained
by Chang et al. \cite{chang1} are shown in Table~II. Clearly, the
estimates are several times larger than the realistic
value. The over estimations possibly indicate a more
decisive role of impurities in FQHE, in contrast to the integral case.
In this context, we observe that Du et al. \cite{du} find that
$m^\ast$ for states with parallel flux attachment is different from
those otherwise. We speculate that while Eq. (\ref{eq14}) is correct in
essence, the right hand side (RHS) will possibly get scaled by such
an effect, compensating for the overestimation of $m^\ast$. The
situation would become clearer only after a proper study is made
with the impurities and other effects put in, as has been done
by Halperin, Lee and Read \cite{halp3} at $T=0$ in vicinity of
$\nu=1/2$.

We now make a few qualitative observations.
Although we are not able to calculate the slope of
the plateau as a function
of temperature, we report that we find a reasonable
agreement between
$\omega_c$ and $E_g$, which is
measured by Du et al. \cite{du} Recall that $\omega_c$
is otherwise an MF artefact.
If we take a rather naive \cite{cage2} view point that
the deviation from the quantization is
proportional to the deviation from the
zero slope, we are then in a position to compare a host of experimental
results with the theoretical prediction.
If we assume that the threshold accuracy for a plateau to be seen is a
minimum of $0.1\%$, we report here that
the temperatures that we estimate are completely
consistent with the temperatures at which these levels have been
seen experimentally.

Finally, Fig.~2 shows the compressibility as a function
of temperature for both the
quantum fluids
(which are incompressible at $T=0$)
for
various values of filling fractions. The generalized expression for the
compressibilty is given by
$\bar{k}=(1/e^2\rho^2) \left[ \Gamma \theta^2 /( \Pi_2\Gamma
+(\Pi_1+\theta)^2 ) \right]$,
which, for IQHE (i.e., for $\theta \rightarrow \infty$ ), reduces to
$\bar{k}=\Gamma /e^2\rho^2$.
The smooth behaviour of $\bar{k}$ with temperature upto a value 5 K
clearly shows that there is no phase transition involving these fluids.
The same
conclusion has been arrived by Chang et al. \cite{chang3} by their
measurement $\rho_{xx} (T)$ for $\nu =2/3$.

To conclude,
We reiterate that what is at the heart
of our analysis is the CS interaction and the MF ansatz. Since all MF
arguments yield an appropriate $\omega_c$
as a natural energy scale, it is
clear that the dependence on $\rho /m^\ast$
at FT must be a common feature
of these models. In particular, Eqs.~(\ref{eq13},\ref{eq14})
for ${\cal R}$ must hold
with the constant on RHS depending on the
particulars of the model. Thus, any experimental
verification of the relations
(\ref{eq13},\ref{eq14}) would shed light on the validity
of the MF ansatz in general.
A precise information of the RHS will hopefully allow us to discern
amongst various models and would complement
measurement of other experimental results

Note: It has recently come to our notice that Zhang has studied
FQHE at finite temperatures with emphasis on a study of the
collective excitations. The preprint \cite{zhang} is
duly referred.

%\newpage

\newpage

\begin{figure}
\caption{The temperatures $T_{{\cal R}}$ as a function of accuracy
	shown for (a)$\nu =1$, (b)$\nu =2/5$ and (c)$\nu =2/3$
	for a typical value of $\rho /m^\ast  =20$ cm$^{-1}$.}
\end{figure}

\begin{figure}
\caption{compressibilities are shown for (a)$\nu =1$,
	(b)$\nu =2/5$ and (c)$\nu =2/3$
	 as function of temperatures
	for a typical value of $\rho /m^\ast  =20$ cm$^{-1}$.}
\end{figure}

\bigskip

\begin{table}
\caption{estimated values of $\rho /m^\ast p$ and hence
  $m^\ast$ from experimental data.}
\begin{tabular}{ccccccc}
Reference & ${\cal R}_0$ & $T$ & $\rho$ & $p$ & estimated & estimated\\
Number & (ppm) & (K) & ($10^{12}$ cm.$^{-2}$) & & $\rho /m^\ast p $
(K) & $m^\ast (m_e)$ \\
\hline
1 & \dec 5.0 & \dec 1.8 & - & - & \dec 9.39 & - \\
21 & \dec 0.2 & \dec 0.5 & 1.0 & 4 & \dec 3.15 & 0.7 \\
22 & \dec 4.2 & \dec 3.0 & - & - & \dec 15.83 & - \\
 22 & \dec 0.017 & \dec 1.2 & - & - & \dec 8.55 & - \\
\end {tabular}
\end{table}

\bigskip

\begin{table}
\caption{estimated values of $m^\ast$ for different filling
  fractions from experimental data of Ref.~24. }
\begin{tabular}{ccccc}
$\nu $ & $(p,s)_\pm $ & ${\cal R}_\pm $ & $\rho $ & estimated \\
 & & & $(10^{11}$ cm.$^{-2}$) & $m^\ast (m_e)$ \\
 \hline
1/3 & $(1,1)_+$ & $3.0\times 10^{-5}$ & 1.53 & 3.89 \\
2/3 & $(2,1)_-$ & $3.0\times 10^{-5}$ & 2.42 & 2.69 \\
2/5 & $(2,1)_+$ & $2.3\times 10^{-4}$ & 2.13 & 3.28 \\
3/5 & $(3,1)_-$ & $1.3\times 10^{-3}$ & 2.13 & 2.54 \\
3/7 & $(3,1)_+$ & $3.3\times 10^{-3}$ & 2.13 & 3.25 \\
\end{tabular}
\end{table}


\begin{references}
\bibitem{fnote} While some plateaus seen experimentally are not
      described by Jain's picture it does seem to be a plausible
      model for a wide class of FQHE states. Further evidence of
      its correctness comes from its application to the $\nu
      =1/2$ state by Halperin Lee and Read. \cite{halp3}
\end{references}
\end{document}